# Brain network modules of meaningful and meaningless objects

Rizkallah J., Benquet P., Wendling F., Khalil M., Mheich A., Dufor O., Hassan M.

*Abstract*— Network modularity is a key feature for efficient information processing in the human brain. This information processing is however dynamic and networks can reconfigure at very short time period (few hundreds of millisecond). This requires neuroimaging techniques with sufficient time resolution. Here we use the dense electroencephalography (EEG) source connectivity methods to identify cortical networks with excellent time resolution (in the order of millisecond). We identify functional networks during picture naming task. Two categories of visual stimuli were presented: meaningful (tools, animals…) and meaningless (scrambled) objects.

In this paper, we report the reconfiguration of brain network modularity for meaningful and meaningless objects. Results showed mainly that networks of meaningful objects were more modular than those of meaningless objects. Networks of the ventral visual pathway were activated in both cases; however a strong occipito-temporal functional connectivity appeared for meaningful object but not for meaningless object. We believe that this approach will give new insights into the dynamic behavior of the brain networks during fast information processing.

## I. Introduction

The brain is organized into structurally and functionally interconnected regions. Brain functions require specific activation of dynamic neural network. In these 'complex' networks, information is continuously processed and integrated between spatially spread but functionally interconnected brain regions with strong temporal dynamics [1]. Common topological features of structural and functional brain networks such as a number of highly connected network hubs 'rich club' have been revealed at micro, meso and macro scale, see [2] for very recent review.

The information processing in the brain is however dynamic and networks can reconfigure at a very short time scale (few hundreds of milliseconds). To track this reconfiguration, neuroimaging techniques with sufficient temporal resolution are required, which is the case of Magnetoencephalography (MEG) and electroencephalography (EEG). In this context, dense EEG (256 electrodes) was used in our work.

A key feature in evaluating the information processing in the brain is its modular structure designs. This modularity was shown to support sustaining wiring cost and creating specialized information [3]. Modularity of brain networks forms how information is circulated and processed: functionally neighboring regions tend to share information are participating in the same module. Thus, modular networks allow for better information processing than nonmodular systems [3].

Methods for detecting modularity usually detect subnetworks that are densely interconnected but externally weakly connected [3], also called network communities which correspond to particular cognitive functions. Modularity was shown to be an essential property of many brain functions such as learning [4] and other various cognitive tasks [5]. Different methods have been developed to compute network modularity [6-10]. The modularity maximization is the widely applied algorithm aimed at partitioning the network into a number of nonoverlapping communities [4].

Here, we track the reconfiguration of brain network modularity during picture naming task including different categories of visual stimuli: meaningful vs. scrambled objects, using the modularity maximization index.

## II. Materials and Methods

### A. EEG recordings and pre-processing

Twenty healthy volunteers with no neurological disease were involved in this study. 80 meaningful and 40 scrambled pictures were displayed and the participants were asked to name them. All pictures were shown as black drawings on a white background. Order of presentation was randomized across participants. Errors in naming were discarded for the subsequent analysis and the signals of one participant were eliminated as data were very noisy.

The brain activity was recorded using dense-EEG, 256 electrodes, system (EGI, Electrical Geodesic Inc.). EEG signals were collected with a 1 kHz sampling frequency providing a high temporal resolution and band-pass filtered

This work was supported by the Rennes University Hospital (COREC Project named BrainGraph, 2015-17). The work has also received a French government support granted to the CominLabs excellence laboratory and managed by the National Research Agency in the "Investing for the Future" program under reference ANR-10-LABX-07-01. This work was also supported by the European Research Council under the European Union's Seventh Framework Programme (FP7/2007-2013) / ERC grant agreement n° 290901.

Rizkallah J. is with Lebanese University and the Laboratoire Traitement du Signal et de l'Image (LTSI), UMR Inserm, Université de Rennes 1, France (corresponding author: Tel: +33 2 23 23 56 05; fax: +33 2 23 23 69 17; e-mail: jennifer.rizkallah.jr@gmail.com).

Benquet P., Mheich A, Wendling F., Hassan M. are with the Laboratoire Traitement du Signal et de l'Image (LTSI), UMR Inserm, Université de Rennes 1, France (e-mail: pbenquet@univ-rennes1.fr, ahmad.mheich@univ-rennes1.fr, fabrice.wendling@univ-rennes1.fr, mahmoud.hassan@univ-rennes1.fr).

Khalil M. is with Lebanese University (e-mail: mohamad.khalil@ul.edu.lb).

Dufor O is with Télécom Bretagne (Institut Mines-Télécom), UMR CNRS Lab-STICC, Brest, France (e-mail: olivier.dufor@telecom-bretagne.eu).

between 3 and 45Hz. Each trial was visually inspected, and epochs contaminated by eye blinking, movements or any other noise source were rejected and excluded from the analysis performed using the EEGLAB [11] and Brainstorm [12] open source toolboxes.

*B. Functional network*

Functional connectivity matrices were calculated using method called 'EEG source connectivity' [13, 14]. It includes two main steps: i) solving the EEG inverse problem to reconstruct regional time series and ii) measuring the statistical couplings, functional connectivity, between these reconstructed regional time series. The weighted Minimum Norm Estimate (wMNE) was used to reconstruct the cortical sources. The functional connectivity was then computed using the phase locking value (PLV) method. This measure (range between 0 and 1) reflects interactions between oscillatory signals by quantifying their phase relationships. The PLVs were estimated at the gamma frequency band (30-45 Hz). The whole brain was parcellated into a set of 68 regions of interest identified by Desikan atlas [15].

*C. Modularity analysis*

Modularity maximization method was proposed to partition the network into nonoverlapping modules by maximizing the modularity index Q [7] defined as:

$$Q = \frac{1}{2m} \sum_{ij} [A_{ij} - \gamma P_{ij}] \delta(\sigma_i, \sigma_j)$$

where $A_{ij}$ represents the weight of the edge between nodes $i$ and $j$ assigned to communities $\sigma_i$ and $\sigma_j$ respectively. $\gamma$ is a structural resolution parameter. $P_{ij} = \frac{K_i K_j}{2m}$ is the expected weight of the edge connecting node $i$ and node $j$ under a specified null model where $K_i = \sum_j A_{ij}$ is the sum of the weights of the edges attached to vertex $i$. The δ-function δ(u, v) is 1 if u = v and 0 otherwise and $m = \frac{1}{2} \sum_{ij} A_{ij}$.

Q measures the difference between the observed connectivity within modules and its expected value for a random graph with the same degree sequence. Thus, a partition achieves a greater value of the modularity index Q closer to unity if the communities are more internally dense than would be expected by chance [16].

*D. Practical issues*

The modularity maximization method was computed to every single matrix 100 times as Q may vary from run to run, due to heuristic property of the algorithm. The value of the modularity index Q was averaged over these 100 runs for each matrix. We compared the results of the Q values computed for the two categories using Wilcoxon rank sum test (with p < 0.01 as significance level).

We also constructed randomized networks with the same strength distribution as the true brain weighted networks, called null model, to determine whether the value of Q was greater or less than expected by chance. This null model was obtained by rewiring the edges of the real brain network N times (here we used N=5). This procedure was repeated 100 times for each real brain network and Louvain modularity maximization was applied also to these randomized networks.

To solve the problem of the optimal community affiliation, we compute co-occurrence matrix, also called 'co-classification matrix' [17]. For each matrix, we calculate the ratio of each node to be in the same module with the other nodes among these 100 runs. This produces a 68x68 matrix where values represent the probability of each two nodes to be in the same module for all runs.

To investigate the consistency of the modules over time, we computed a second co-occurrence matrix by calculating the ratio of each node to be with the other nodes in the same module for a given time window. The modularity was computed using the Brain Connectivity Toolbox (BCT) [18] and the networks were visualized using EEGNET [19].

## III. RESULTS

*A. Modularity index*

We investigated the network organization of the brain during the cognitive task over the entire task using the modularity index as a measure of the amount of network modularity. We found that the Q values of the real network (0.12 < $Q_{real\_network}$ < 0.14) was significantly larger than expected in a random network (0.01 < $Q_{null\_model}$ < 0.03). The modularity index Q was significantly different between the two categories (scrambled vs. meaningful) at different time scales. Here we show only the results for the longest time period (177ms: 260ms) showing the significant difference (Wilcoxon, p < 0.001) (Fig. 1).

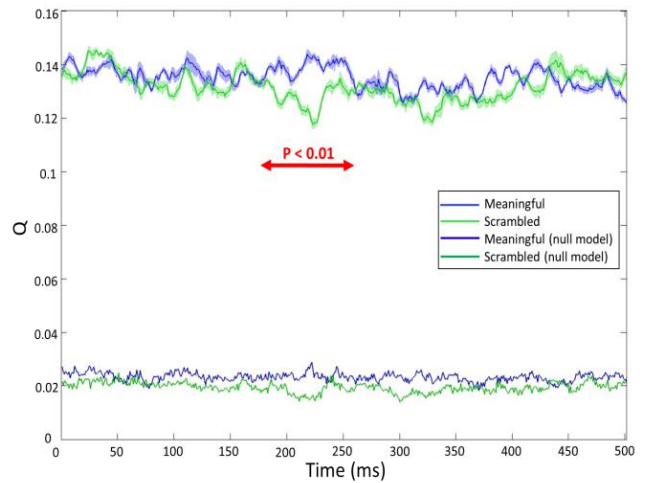

Figure 1. Evolution of the modularity index Q values with time for each category compared to null models. Horizontal red line represents the longest window with significant difference between the two categories.

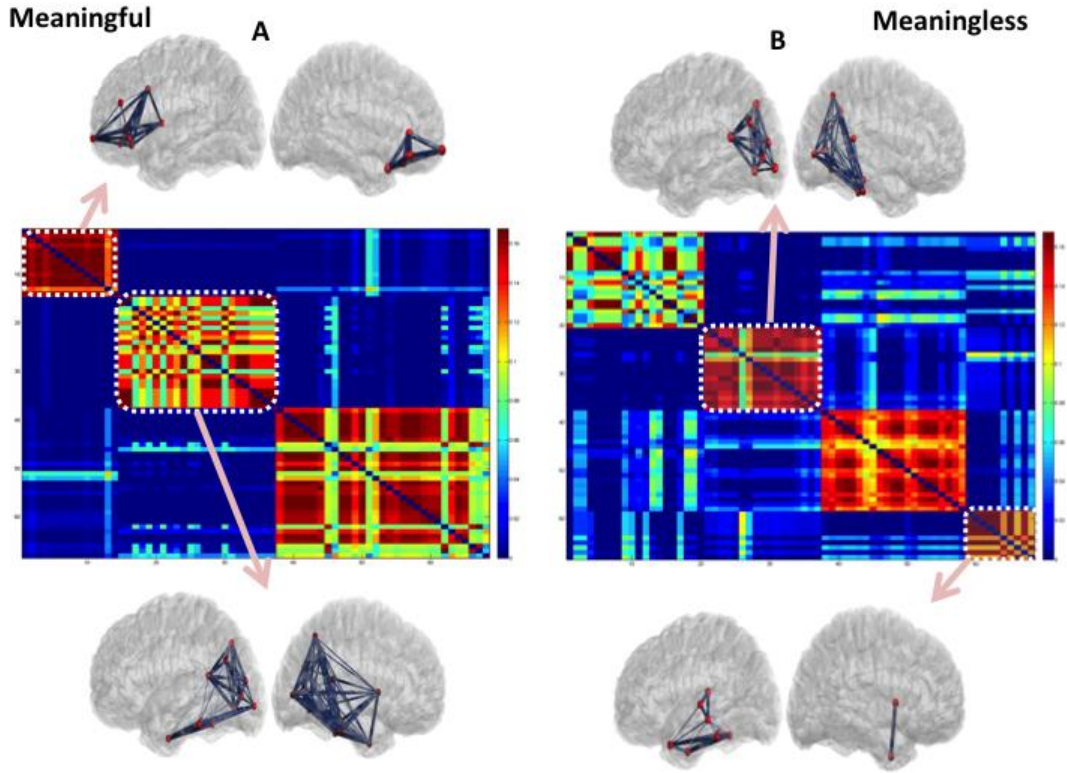

Figure 2. Co-occurrence matrices (over runs and time period) of networks associated to meaningful (A) and scrambled (B) objects. Axes represent the brain regions (from 1 to 68). The two modules with highest ratio are highlighted in dashed white boxes and the corresponding networks are also presented. The size of the edges represents the co-occurrence ratio.

*B. Network segregation*

We then computed the co-occurrence matrices at this time period. For the meaningful objects, network was segregated into 3 modules (Fig. 2A) while the network of the scrambled pictures was partitioned into 4 modules (Fig. 2B).

In figure 2, we show also the networks corresponding to the identified modules. We showed the two modules with the highest occurrence ratio at the analyzed period. For the meaningful network (Fig. 2A), the first module was mainly located in the frontal lobe, the second module was found principally in the occipital and temporal lobes. For the scrambled network (Fig. 2B), the first module involved mainly the occipital regions and the second module was mainly involving a focal left temporal network.

Note that scrambled image generate separated occipital and temporal modules without functional connectivity within the occipito-temporal pathway. Inversely meaningful image induce functional connectivity between the primary visual cortex, lateral occipital cortex, temporal inferior cortex and the temporal pole.

## IV. DISCUSSION AND CONCLUSION

Here we reported the results of characterizing the modular reconfiguration of functional brain networks during a task of naming meaningful and meaningless objects. We used EEG source connectivity method to identify functional networks at the cortical level from scalp EEG recordings. This method allowed us to track network at few hundreds millisecond time scale. Modularity index was used to track network reconfiguration. Our results showed a significant difference in brain network modularity when processing meaningless or meaningful objects.

Results showed that the brain network was partitioned into 3 modules for the meaningful network and 4 modules for the scrambled network. These results showed that the brain is more segregated (higher modularity) when naming meaningful objects than meaningless objects. Occipital module (related to the visual information processing) was observed in both conditions with an implication of the temporal lobe in the case of meaningful objects. This temporal lobe is well known to be related to the semantic processing [20]. Frontal module was observed in the case of meaningful objects. A module that involves solely the temporal regions was observed only in the case of the scrambled objects.

Here we focused in the modules occurrence during one of the time windows showing significant difference (in modularity index Q) between the two categories. However, we did not investigate the dynamics of the modules reconfigurations (which module occur before other modules or if two modules are occurred at the same time). This dynamic aspect is the main objective of our ongoing work. In addition, the study was performed on averaged connectivity matrices, a subject-specific analysis will be also considered.

Regarding the technical issues, we considered several values of the resolution parameter γ to choose the best one for our experiment, γ=1 was the good compromise between number for modules and nodes associated with each module (as reported also by many studies [4]). Louvain community detection method was applied 100 times as the values of Q change from a run to another. The solution was to compute a matrix of the probability of each two nodes to be in the same module for all runs. The resultant matrix was called 'co-occurrence' matrix. This procedure was also applied in the time domain to keep only the modules that resist in time and present in the entire time window.

According to the "communication through coherence" theory [21], the brain dynamically associate and process information in distinct cortical areas via synchronized oscillations. This synchronization (associated with phase, frequency, duration and spatiotemporal features) between two neuronal groups produces rhythmic excitability variations that create temporal windows for communication. It was shown that synchronization between gamma oscillations represent essential process in cortical computation [22].

In this context, tracking brain oscillations at the level of cortical sources using dense-EEG source connectivity method in the gamma frequency band during a picture naming task provide appropriate conditions in order to detect such communication phenomena. Our results showed that functional connectivity appeared between the occipital and the inferior temporal cortex for meaningful objects but not for scrambled. These results can be explained by the fact that resonance between brain areas of visual detection and brain areas of categorization was only associated with meaningful objects.

Finally, here we show a new way of characterizing differences between the processing of meaningful and meaningless objects in the brain. The approach is based on the modularity of functional brain networks at very short time scale. We believe that this approach will give new insights into the dynamic behavior of the brain during information processing. Our approach might help to understand how connectivity, within and beyond the ventral visual pathway, dynamically changes over time during object recognition.

REFERENCES

[1] O. Sporns, G. Tononi, and G. M. Edelman, "Connectivity and complexity: the relationship between neuroanatomy and brain dynamics," *Neural Networks*, vol. 13, pp. 909-922, 2000.

[2] O. Sporns, "Connectome Networks: From Cells to Systems," in *Micro-, Meso-and Macro-Connectomics of the Brain*, ed: Springer, 2016, pp. 107-127.

[3] O. Sporns and R. F. Betzel, "Modular brain networks," *Annual review of psychology*, vol. 67, pp. 613-640, 2016.

[4] D. S. Bassett, N. F. Wymbs, M. A. Porter, P. J. Mucha, J. M. Carlson, and S. T. Grafton, "Dynamic reconfiguration of human brain networks during learning," *Proceedings of the National Academy of Sciences*, vol. 108, pp. 7641-7646, 2011.

[5] M. A. Bertolero, B. T. Yeo, and M. D'Esposito, "The modular and integrative functional architecture of the human brain," *Proceedings of the National Academy of Sciences*, vol. 112, pp. E6798-E6807, 2015.

[6] M. E. Newman, "Modularity and community structure in networks," *Proceedings of the national academy of sciences*, vol. 103, pp. 8577-8582, 2006.

[7] V. D. Blondel, J.-L. Guillaume, R. Lambiotte, and E. Lefebvre, "Fast unfolding of communities in large networks," *Journal of statistical mechanics: theory and experiment*, vol. 2008, p. P10008, 2008.

[8] T. Hastie, R. Tibshirani, and J. Friedman, "Unsupervised learning," in *The elements of statistical learning*, ed: Springer, 2009, pp. 485-585.

[9] M. Rosvall and C. T. Bergstrom, "Maps of random walks on complex networks reveal community structure," *Proceedings of the National Academy of Sciences*, vol. 105, pp. 1118-1123, 2008.

[10] M. B. Hastings, "Community detection as an inference problem," *Physical Review E*, vol. 74, p. 035102, 2006.

[11] A. Delorme and S. Makeig, "EEGLAB: an open source toolbox for analysis of single-trial EEG dynamics including independent component analysis," *Journal of neuroscience methods*, vol. 134, pp. 9-21, 2004.

[12] F. Tadel, S. Baillet, J. C. Mosher, D. Pantazis, and R. M. Leahy, "Brainstorm: a user-friendly application for MEG/EEG analysis," *Computational intelligence and neuroscience*, vol. 2011, p. 8, 2011.

[13] M. Hassan, O. Dufor, I. Merlet, C. Berrou, and F. Wendling, "EEG source connectivity analysis: from dense array recordings to brain networks," *PloS one*, vol. 9, p. e105041, 2014.

[14] M. Hassan, P. Benquet, A. Biraben, C. Berrou, O. Dufor, and F. Wendling, "Dynamic reorganization of functional brain networks during picture naming," *Cortex*, vol. 73, pp. 276-288, 2015.

[15] R. S. Desikan, F. Ségonne, B. Fischl, B. T. Quinn, B. C. Dickerson, D. Blacker*, et al.*, "An automated labeling system for subdividing the human cerebral cortex on MRI scans into gyral based regions of interest," *Neuroimage*, vol. 31, pp. 968-980, 2006.

[16] B. H. Good, Y.-A. de Montjoye, and A. Clauset, "Performance of modularity maximization in practical contexts," *Physical Review E*, vol. 81, p. 046106, 2010.

[17] A. Fornito, A. Zalesky, and E. Bullmore, "Fundamentals of brain network analysis," 2016.

[18] M. Rubinov and O. Sporns, "Complex network measures of brain connectivity: uses and interpretations," *Neuroimage*, vol. 52, pp. 1059-1069, 2010.

[19] M. Hassan, M. Shamas, M. Khalil, W. El Falou, and F. Wendling, "EEGNET: An Open Source Tool for Analyzing and Visualizing M/EEG Connectome," *PloS one*, vol. 10, p. e0138297, 2015.

[20] A. Clarke and L. K. Tyler, "Understanding what we see: how we derive meaning from vision," *Trends in cognitive sciences*, vol. 19, pp. 677-687, 2015.

[21] P. Fries, "Rhythms for cognition: communication through coherence," *Neuron*, vol. 88, pp. 220-235, 2015.

[22] P. Fries, "Neuronal gamma-band synchronization as a fundamental process in cortical computation," *Annual review of neuroscience*, vol. 32, pp. 209-224, 2009.